\documentclass{jetpl}
\twocolumn
\usepackage{graphicx}
\usepackage{graphics}
\usepackage{psfrag}
\lat

\title{Spin crossover: the quantum phase transition induced by high pressure}

\rtitle{Spin crossover: the quantum phase transition \ldots}

\sodtitle{Spin crossover: the quantum phase transition induced by high pressure}
\author{A.\,I.\,Nesterov$^{+}$\/\thanks{e-mail: nesterov@cencar.udg.mx},
S.\,G.\,Ovchinnikov$^{{-}{**}}$\/\thanks{e-mail: sgo@iph.krasn.ru}}

\rauthor{A.\,I.\,Nesterov, S.\,G.\,Ovchinnikov}

\sodauthor{Nesterov, Ovchinnikov}

\address{$^+$ CUCEI, Universidad de Guadalajara, Guadalajara, M\'exico~\\
$^-$ L.V. Kirensky Institute of Physics, SBRAS, Krasnoyarsk, Russia\\
$^{**}Siberian Federal University, 660062, Krasnoyarsk, Russia$}

\date{\today}

\abstract{The relationship is established between the Berry phase and spin crossover in condensed matter physics induced by high pressure. It is shown that the geometric phase has topological origin and can be considered as the order parameter for such transition. }


\begin{document}
\bibliographystyle{aipprocl}

\maketitle

Spin crossover in condensed matter physics is a transformation of a system with one spin $S_1$ at each lattice site into another state with spin $S_2$ induced by some external parameter like strong magnetic field, high pressure etc. It accompanies by the energy level $E_1$ and $E_2$ crossing, where $E_a$ is the local energy of the magnetic ion with spin $S_a$ ($a=1,2$). Recently spin crossovers in magnetic oxides have been found under high pressure in  $FeBO_3$ \cite{STL},   $CdFe_3$ \cite{GKL},   $BiFeO_3$  \cite{GSL},  $Fe_3O_4$ \cite{DHO}. Below the Curie temperature of magnetic order spin crossover is accompanied by the sharp change of the magnetization, nevertheless it may be observed in the paramagnetic state like in $CdFe_3(BO_3)_4$ \cite{GKL} as the sharp change of the XES satellite/main peak intensity ratio with pressure increase. There is no thermodynamic order parameter that can distinguish one phase versus the other. In this paper we discuss the low temperature limit and claim that at $T=0$ spin crossover is a quantum phase transition. The order parameter for such transition has topological origin and we calculate the geometrical phase that characterizes the spin crossover.

Quantum phase transition (QPT) is characterized by qualitative changes of the ground state of many body system and occur at the zero temperature \cite{SS}. QPT being purely quantum phenomena driven by quantum fluctuations, is associated with levels crossing and imply the lost analyticity in the energy spectrum. In the parameter space the points of non-analyticity, being referred to as critical points, define the QPT \cite{SS}. For the Hermitian Hamiltonian coalescence of eigenvalues results in different eigenvectors, and related degeneracy referred to as `conical intersection' is known also as `diabolic point' \cite{B,H1}.

Since QPT is accomplished by changing some parameter in the Hamiltonian of the system, but not the temperature, its description in the standard framework of the Landau-Ginzburg theory of phase transitions failed, and identification of an order parameter is still an open problem \cite{SGC}. In this connection, an issue of a great interest is recently established  relationship between geometric phases and quantum phase transitions \cite{PC,CP,Zhu,HA}. This relation is expected since the geometric phase associated with the energy levels crossings has a peculiar behavior near the degeneracy point. It is supposed that the geometric phase, being  a measure of the curvature of the Hilbert space, is able to capture drastic changes in the properties of the ground states in presence of QPT \cite{CP,Zhu,HA,ZSL}

In the context of the Berry phase the diabolic point is associated with `fictitious magnetic monopole' as follows. Assume that for adiabatic driving quantum system two energy levels may cross. Then the energy surfaces form the sheets of a double cone, and its apex is called a ``diabolic point'' \cite{BW}. Since for generic Hermitian Hamiltonian the codimension of the diabolic point is three, it can be characterized by three parameters $\mathbf R= (X,Y,Z)$. The eigenstates $|n,\mathbf R \rangle$ give rise to the Berry's connection defined by ${\mathbf A}_n(\mathbf R)= i\langle n,\mathbf R| \nabla_{\mathbf R} |n,\mathbf R \rangle$, and the curvature
$\mathbf B_n = \nabla_{\mathbf R} \times {\mathbf A}_n $ associated
with ${\mathbf A}_n$ is the field strength  of `magnetic' monopole located at
the diabolic point \cite{B0,BD}. The Berry phase $\gamma_n= \oint_{\mathcal C }{\mathbf A}_n \cdot d \mathbf R$ is interpreted as a holonomy associated with the parallel transport along a circuit $\mathcal C$ \cite{SB}.

{\em  Geometric phases and quantum phase transitions. --} Consider the
diagonalizable Hamiltonian $H(\lambda)= \sum_{i=1} E_i (\lambda)|\psi_i(\lambda)\rangle \langle \psi_i(\lambda)|$, depending on the parameters $\lambda^a$, $a=1,2,\dots,r$. Its ground state is given by $|\psi_g (\lambda)\rangle = \otimes^N_{i=1}|\psi_i (\lambda) \rangle$, and employing the standard formula for computing of the Berry phase, we obtain
\begin{align}\label{B0}
\gamma = i\oint_{\cal C} \langle \tilde \psi_g(\lambda)| \frac{\partial}{\partial \lambda^a} |\psi_g(\lambda)\rangle d \lambda^a = \sum^N_{i=1} \gamma_i
\end{align}
where $\gamma_i =\oint_{\cal C}dA^{(i)}$ is the geometric phase associated with the eigenvector $|\psi_i(\lambda) \rangle$. Then applying the Stokes theorem we obtain
\begin{align}
\gamma = -i\sum^N_{i=1} \sum^N_{m\neq i}\iint_{\Sigma}\frac {\langle   \psi_i|\nabla_a H |\psi_m\rangle
 \langle   \psi_m| \nabla_b H |\psi_i\rangle d\lambda^a\wedge d\lambda^b } {(E_m-E_i)^2}
 \nonumber
 \end{align}
It follows herefrom that the curvature $F^{(i)}= dA^{(i)}$ diverges at the degeneracy points, where the energy levels, say $E_n$ and $E_{n+1}$, are crossing, and reaches its maximum values at the avoided level crossing points. Thus, the critical behavior of the system is reflected in the geometry of the Hilbert space through the geometric phase of the ground state.

Since in the vicinity of the level crossing point only the two-dimensional Jordan block related to the level crossing makes the most considerable contribution to the quantum evolution, the $N$-dimensional problem can be described by the effective two-dimensional Hamiltonian which can be obtained as follows. Let $\lambda_c$ be a crossover point at which the energies $E_n(\lambda_c)$ and $E_{n+1}(\lambda_c)$ coalescenc. In the two-dimensional subspace corresponding to $E_n(\lambda_c)$ and $E_{n+1}(\lambda_c)$, we choose an orthonormal basis  $\{|0\rangle,|1\rangle\}$ and complement it to the complete basis of the $N$-dimensional Hilbert space by adding the eigenvectors  $|\psi_k (\lambda_c) \rangle$ $(k\neq n,n+1)$.

Now, an arbitrary state $|\psi(t)\rangle$ can be expanded as $|\psi(t)\rangle = \alpha(t) |0\rangle + \beta (t)|1\rangle + \sum^{N-1}_{k\neq k\neq n,n+1}c_k(t)|\psi_k (\lambda_c) \rangle$. Inserting this expansion into the time-dependent Schr\"odinger equation, we obtain the coefficients $\alpha$ and $\beta$ as the solution of the two-dimensional Schr\"odinger equation
\begin{align}\label{Sch2}
i \frac{\partial }{\partial t}|u(t)\rangle = \mathcal H_{ef}(\lambda)|u(t)\rangle,
\end{align}
where
\begin{equation}\label{H1}
 \mathcal H_{ef}(\lambda) =\left(
  \begin{array}{cc}
    \lambda_0 + Z & X-iY \\
   X+iY & \lambda_0 - Z\\
  \end{array}
\right)
\end{equation}
and $|u(t)\rangle = \bigg (
  \begin{array}{c}
    \alpha \\
    \beta \\
  \end{array}\bigg)$. The matrix elements in Eq. (\ref{H1}) are determined by
\begin{align}
\lambda_0 = \frac{1}{2}(\langle 0| H(\lambda)|0\rangle + \langle 1|  H(\lambda)|1\rangle) \label{R1a},\\
X= \frac{1}{2}(\langle 0| H(\lambda)|1\rangle + \langle 1|  H(\lambda)|0\rangle), \\
Y= \frac{i}{2}(\langle 0|  H(\lambda)|1\rangle - \langle 1|  H(\lambda)|0\rangle ),\\
Z= \frac{1}{2}(\langle 0|  H(\lambda)|0\rangle- \langle 1|  H(\lambda)|1 \rangle )\label{R1b}.
\end{align}

Thus, in the neighborhood of diabolic point only terms related to the invariant subspace formed by the corresponding two-dimensional Jordan block make substantial contributions. The $N$-dimensional problem becomes effectively two-dimensional, and the quantum system can be described by the effective two-dimensional Hamiltonian $H_{ef}= \lambda_0 {1\hspace{-.125cm}1}+\mathbf R\cdot \boldsymbol \sigma$, where $\mathbf R(\lambda)= (X,Y,Z) $ (for details see \cite{Arn,KMS,NAI}).

The geometric phase in neighborhood of the diabolic point can be written as follows
\begin{align}\label{B7}
\gamma \approx \frac{1}{2} \int_{\Sigma} \frac{\mathbf R \cdot d\mathbf S }{R^3} + \sum_{i\neq n,n+1} \gamma_i(\mathbf R )
\end{align}
where integration is performed over the surface $\Sigma \subset S^2$. The behavior of the geometric phase described by the first term is independent of a peculiarities of quantum-mechanical system. One can consider the Bloch sphere as an universal parameter space for description of QPT in the vicinity of the critical point \cite{NO}.

Following \cite{CP}, we define the overall geometric phase of the ground state as $\gamma_g = (1/N)\sum^N_{i=1} \gamma_i$. In the thermodynamical limit $\gamma_g = \int \gamma(x)d\mu(x)$, where $d\mu(x)$ is the suitable measure.
As has been shown by Zhu \cite{Zhu} on example of $XY$ spin chain, the overall geometric phase associated with the ground state exhibits universality, or scaling behavior in the vicinity of the critical point. In addition, the geometric phase allows to detect the critical point in the parameter space of the Hamiltonian \cite{PC,CP,HA,Zhu,ZSL}. These works indicate that the overall geometric phase $\gamma_g$ can be considered as the universal order parameter for description of QPT.

{\em The model. --} The multielectron ion with in the crystal field has the energies of terms for $d^n$  configurations determined numerically by the Tanabe-Sugano diagrams \cite{TS} as a solution of the eigenvalue problem. Simple analytical calculations of the low energy terms with different spin value that is sufficient to study spin crossover has been done recently \cite{SGO}. The crystal field parameter $\Delta$ increase linearly with pressure $P$. Thus the multielectron energies for spin  $S_1$  and  $S_2$  ( $E_1$  and  $E_2$ ) are also linear functions of $P$. To distinguish two different spin states in the lattice we introduce the Ising  pseudospin states $|i\rangle$  and  $|-i\rangle$   for  $|d_i^n,S^i_1\rangle$ and $|d_i^n,S^i_2\rangle$, where $i$ runs over all sites in the lattice. Thus we neglect the spin degeneracy of the $d_i^n$ terms but capture the possibility of energy level crossing that is the essential part of the spin crossover. Then, in the basis  $|+i\rangle$, $|-i\rangle$, the Hamiltonian of the system can be written as follows
\begin{eqnarray}\label{H1a}
H= \sum_i\big({\lambda^i_0}{1\hspace{-.125cm}1} +{\varepsilon_i}\hat\sigma^z_{i}) + \sum_{ij}H_{ij},
\end{eqnarray}
where $\lambda^i_0=(E^i_1+  E^i_2)/2$, $\varepsilon_i =(E^i_1 -  E^i_2)/2 $, and  ${1\hspace{-.125cm}1} $, $\hat\sigma_z$ are the identity and Pauli matrices, respectively; the Hamiltonian of interaction between the spins being $H_{ij}$.
The main contribution to the $H_{ij}$ is given by the Heisenberg ecxange interaction. We consider spin crossover far from the thermodynamic phase transition in the paramagnetic phase, it allows us to simplify this interaction and substitute it with the effective mean field. This mean field is spatially uniform for the ferromagnetic insulator or two-sublattice for the antiferromagnet one. Examples given above  \cite{STL,GKL,GSL,DHO} correspond to the anti- or ferrimanetics. In any case this mean field just renormalizes the interionic multielectron energies $E_1$ and $E_2$. Nevertheless this mean field results in the collective behaviour of spin system in the crystal, that is why we can treat all spins in the space uniform states. Thus spin crossover at zero temperature is the transition from one spin ordered state to another spin ordered state. Another interaction that is smaller then the exchange one is given by relativistic anisotropy contribution to the $H_{ij}$. For example a spin-orbital interaction can mix different spin states inside single ion, and it occurs to be important in our problem.

In what follows we will consider the simplified spatially uniform model described by the following Hamiltonian
\begin{eqnarray}\label{H2}
H = \sum^N_{i=1}(\lambda_0 {1\hspace{-.125cm}1}+ \varepsilon\hat\sigma_z + \lambda\hat\sigma_{+} + \lambda^*\hat\sigma_{-}).
\end{eqnarray}
The spin-orbit coupling $\lambda$ mixes the different spin states, and it plays the role of quantum fluctuations in our Ising pseudospin basis. Both $\lambda_0$  and $\varepsilon$ are pressure dependent. Further we assume $\varepsilon(P) = \varepsilon_0 -aP$. The crossover takes place when $\varepsilon(P_c) = 0$ at $P=P_c=\varepsilon_0/a$.

\begin{figure}[tbp]
\begin{center}
\scalebox{0.35}{\includegraphics{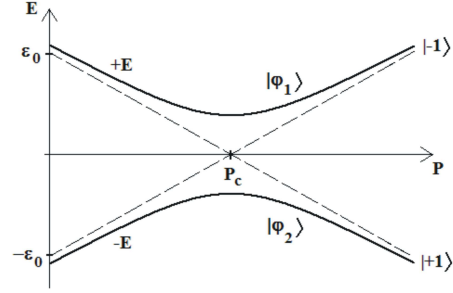}}
\end{center}
\caption{Fig. 1. Energy levels crossing}
\label{E}
\end{figure}

The Hamiltonian (\ref{H2}) is diagonalized by the unitary transformation
\begin{eqnarray}
\label{Eq1}
  |\varphi_1\rangle &=&\frac{1}{2}( u|+1\rangle + v|-1\rangle) \\
  |\varphi_2\rangle &=&\frac{1}{2}(-v|+1\rangle + u|-1\rangle)
\end{eqnarray}
where $|+1\rangle$ and $|-1\rangle$ are eigenstates of the operator $\hat\sigma_z$:  $\hat\sigma_z |\pm 1\rangle = \pm|\pm 1\rangle$;  $u=\sqrt{1+\varepsilon/E}$, $v=\sqrt{1-\varepsilon/E}$ and $E=\sqrt{\varepsilon^2 + \rho^2}$, we denote $\rho = |\lambda|$. After diagonalization we obtain $H= \sum H_i$, where
\begin{eqnarray}
H_i = \varepsilon_+|\varphi_1\rangle\langle\varphi_1| + \varepsilon_-|\varphi_2\rangle\langle\varphi_2|
\end{eqnarray}
and $\varepsilon_{\pm} = \lambda_0 \pm E$. Due to perturbation there is a finite gap $2\rho$  between eigenstates at the crossover point $\varepsilon =0$ (see Fig.\ref{E}). At $\lambda \rightarrow 0$, $u \rightarrow \sqrt{1+ \varepsilon/|\varepsilon|}$ and $v \rightarrow \sqrt{1- \varepsilon/|\varepsilon|}$ . When $P < P_c$ we have $u \rightarrow \sqrt{2}$, $v \rightarrow 0$ and $|\varphi_1\rangle \rightarrow |+1\rangle$, $|\varphi_2\rangle \rightarrow |-1\rangle$. After crossover ($P>P_c$) $u \rightarrow 0$, $v \rightarrow \sqrt{2}$ and $|\varphi_1\rangle \rightarrow |-1\rangle$, $|\varphi_2\rangle \rightarrow |+1\rangle$. For $\lambda \neq 0$ we can ascribe the definite spin to the ground term $|\varphi_2\rangle$ only asymptotically.
In order to study the geometric phase in this system,  we introduce a new  Hamiltonian $\cal H(P,\lambda,\varphi) =U(\varphi) H U^{\dagger}(\varphi)$, where $U(\varphi) =  e^{i\frac{\varphi}{2} \sigma_z}$ and $\varphi$ is slowly varying parameter, $0 \leq \varphi < 2\pi$ \cite{NO}. The transformed Hamiltonian $H_i$ takes the form
\begin{eqnarray}\label{eqH2a}
{\cal H}_i= \left(
     \begin{array}{cc}
       \lambda_0 & 0 \\
       0 & \lambda_0 \\
     \end{array}
   \right)
   + \left(
       \begin{array}{cc}
         \varepsilon &  \lambda^\ast  e^{-i\varphi}\\
         \lambda  e^{i\varphi} & -\varepsilon\\
       \end{array}
     \right)
  \end{eqnarray}
The energy spectrum is given by $\varepsilon_{\pm}= \lambda_0 \pm \sqrt{\varepsilon^2 + \rho^2}$, and the ground state energy is $ \varepsilon_{-}$. The instantaneous eigenvectors are found to be
\begin{eqnarray}
&|u_{-}\rangle = \left(\begin{array}{c}
-e^{-i\varphi}\sin\frac{\theta}{2}\\
\cos \frac{\theta}{2} \end{array} \right ), \quad
&|u_{+}\rangle = \left(\begin{array}{c}
                  \cos\frac{\theta}{2}\\
                  e^{i\varphi}\sin\frac{\theta}{2}
                  \end{array}\right) \label{r}
\end{eqnarray}
where $\cos\theta = \varepsilon/\sqrt{\varepsilon^2 + \rho^2} $. Coupling of eigenvalues $\varepsilon_{+}$ and $\varepsilon_{-}$ occurs at the diabolic point located at the origin of coordinates.

\begin{figure}[tbh]
\begin{center}
\scalebox{0.6}{\includegraphics{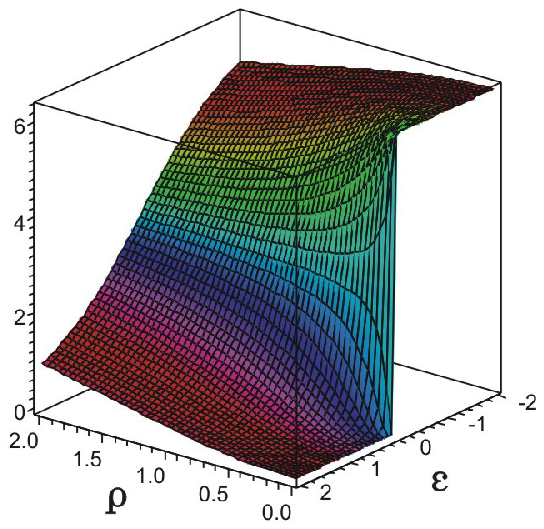}}
\scalebox{0.6} {\includegraphics{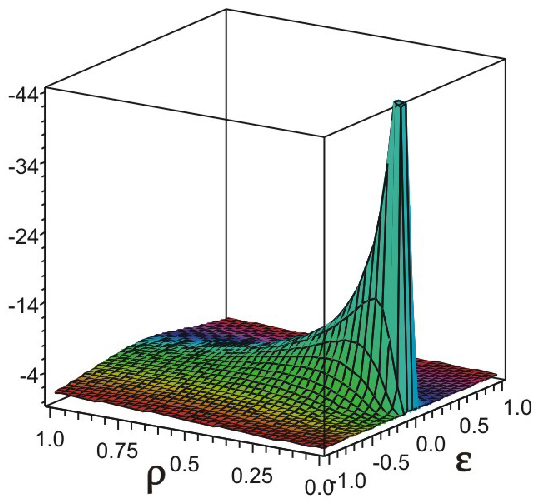}}
\end{center}
\caption{Fig. 2. Geometric phase $\gamma$ (left) and its derivative  $\partial \gamma/\partial\varepsilon$ (right) as a function of the Hamiltonian
parameters $\rho$ and $\varepsilon$. There is clear step-function behavior at the diabolic point $\rho=\varepsilon=0$. }
\label{DP1}
\end{figure}

The connection one-form associated with the ground state is given by
 $ A= \langle u_{-}|d| u_{-}\rangle= \frac{1}{2}(1 -\cos\theta)d\varphi$,
and computation of geometric phase yields
$\gamma = \oint_{\mathcal C} A,$
where integration is performed over the contour $\mathcal C$ on the two-dimensional sphere $S^2$. Let us assume that the contour $\mathcal C$ of integration  is chosen as $\theta= \rm const$. Then the geometric phase
related to the ground state is
\begin{equation}\label{Q2}
\gamma = \pi(1 - \cos\theta) = \pi\left (1- \frac{\varepsilon}{\sqrt{\varepsilon^2 + \rho^2}}\right ).
\end{equation}
The lost of analyticity occurs at the diabolic point located at the origin of the parameter space $(\Re\lambda,\Im\lambda,\varepsilon)$. In vicinity of the diabolic point the geometric phase behaves as a step function
\begin{eqnarray}\label{geometric phase1a}
 \gamma= \Bigg \{
\begin{array}{l}
 0, \;{\rm for}\; \rho =0, \; \varepsilon \rightarrow + 0\; (\theta \rightarrow  0) \\
 2 \pi, \;{\rm for}\; \rho =0, \; \varepsilon \rightarrow -0 \; (\theta \rightarrow \pi)
 \end{array}
\end{eqnarray}
The geometric phase $\gamma$ and its derivative $\partial \gamma/\partial\varepsilon$ versus $\rho, \varepsilon$  are plotted in Fig. \ref{DP1}. As can be observed, the geometric phase behaves as the step-function near the diabolic point, and at the diabolic point one has the discontinuity of the geometric phase with the gap of $2\pi$.

The overall geometric phase $\gamma_g= (\pi/N)\sum_{i}\gamma_i$ can be written as $\gamma_g = \pi (1+ \partial E_g/\partial \varepsilon)$, where $E_g$ is the ground state energy per spin \cite{NO}. Besides, one can show that $\gamma_g = \pi(1 +  \langle \hat\sigma_z\rangle)$, where $\langle \hat\sigma_z\rangle = (1/N)\langle \psi_g | \hat\sigma_z|\psi_g\rangle  $ is the average sublattice magnetization per ion. In our model $\gamma_g$ coincides with partial geometric phase $\gamma_i= 1- \cos\theta$, therefore the step-like behavior of $\langle \hat\sigma_z\rangle$ due to high spin-low spin term crossover reported in \cite{SGO1} has the topological nature.

{\em Concluding remarks. --} In the limit $\rho \rightarrow 0$ the crossover becomes the QPT. The latter has a pure topological nature and emerges as the quantum transition between the ground states with the distinct winding numbers \cite{VG1,VG}. This quantum number, being defined by the geometric phase, is related to a winding number of the map $S^1 \rightarrow U(1)$. Thus, the geometric phase can be considered as the {\em topological order parameter} in the spin crossovers phenomena.

\end{document}